# PKS 2349−014: A LUMINOUS QUASAR WITH THIN WISPS, A LARGE OFF-CENTER NEBULOSITY, AND A CLOSE COMPANION GALAXY [1]

John N. Bahcall, Sofia Kirhakos

Institute for Advanced Study, School of Natural Sciences, Princeton, NJ 08540

and

Donald P. Schneider

Department of Astronomy and Astrophysics, The Pennsylvania State University, University Park, PA 16802

astro-ph/9504076  21 Apr 95

## ABSTRACT

HST images (WFC2) of PKS 2349−014 show that this luminous nearby quasar is interacting with diffuse (presumably galactic) material. Two thin wisps that have a total extent of about 20 kpc (for $H_0 = 100$ km s$^{-1}$ Mpc$^{-1}$ and $\Omega_0 = 1.0$) are observed to approximately surround the quasar. One of the wisps appears to pass through a companion galaxy that is located at a projected distance of three kpc from the center of the quasar light. The companion galaxy, if located at the distance of PKS 2349−014, has an intrinsic size and luminosity similar to the Large Magellanic Cloud. A faint extended nebulosity, which is detected over a region of 35 kpc by 50 kpc and is centered about 5 kpc from the quasar nucleus, overlaps the wisps. The immediate environment of PKS 2349−014 is different from the enivronments of the other eight luminous quasars that we have studied previously with HST. If the multiple light components of the HST images are fit to a single de Vaucouleurs profile, as was done in previous analyses of ground−based data, then the result obtained for the total luminosity of the model galaxy is in agreement with the earlier ground-based studies.

*Subject headings:* quasars: individual (PKS 2349−014)

# 1. INTRODUCTION

We present Hubble Space Telescope (HST) images of PKS 2349−014 that show that this luminous, nearby quasar is interacting with diffuse—presumably galactic—material. Before proceeding further, the reader is urged to look at the three enclosed figures.

Since the epochal study of galactic interactions by Toomre & Toomre (1972), there have been many suggestions that galactic mergers or tidal interactions may be related to the quasar phenomenon (for recent reviews of this work, see for example the papers in Shlosman 1994, and the summary discussion by Heckman 1989; cf. Gunn 1979; Stockton 1982; the classic image of 3C 249.1 in Stockton and MacKenty 1983; the critical discussion of Stockton 1990; Hutchings & Neff 1992). Numerical simulations have shown that features like the wisps that are evident in Figure 1 are characteristic of gravitational interactions among galaxies. Guided by these previous studies, we suggest that the HST images of PKS 2349−014 provide detailed and dramatic evidence for galactic interactions in a luminous quasar.

In a crude condensation of the HST images, we will speak of PKS 2349−014 in the following discussion as if the quasar were composed of five distinct components: 1) an unresolved (stellar) nucleus; 2) two thin curved wisps; 3) a large, faint nebulosity centered a few arcseconds from the quasar nucleus; 4) a close but physically distinct companion galaxy; and 5) residual light (which could be a host galaxy) centered on the stellar quasar. In practice, it is impossible to separate uniquely the different, but overlapping, components. The top panel of Figure 2 gives a schematic picture of the five components of PKS 2349−014. Table 1 gives a summary of our best estimates of the principal characteristics of these five components.

PKS 2349−014 is one of 20 luminous nearby quasars that are being studied in our imaging program with the Hubble Space Telescope to determine the nature of their environments. The results obtained by analyzing images of the first eight quasars in this sample have been described previously (see Bahcall, Kirhakos, & Schneider 1994, 1995; hereafter Paper I and Paper II). None of the previously-analyzed quasars in our sample showed evidence for interactions similar to what is present in the images of PKS 2349−014. The primary selection criteria for inclusion in our sample are luminosity (high) and redshift (small). All of the quasars in the sample were chosen from the Véron-Cetty & Véron (1993) catalog and have $z \leq 0.30$ and $M_V \leq -22.9$ for $H_0 = 100$ km s$^{-1}$ Mpc$^{-1}$ and $\Omega_0 = 1.0$.

Table 1: Components of PKS 2349−014

| Component | Size | $m_{606W}$ | Comment |
|---|---|---|---|
| Unresolved nucleus | — | 15.3 | — |
| Wisps | $10'' \times 2''$ | 17.9 | Tidal tails? |
| Extended nebulosity | $25'' \times 18''$ | 18.0 | Large, off-center galaxy? |
| Companion galaxy | $1'' \times 0.7''$ | 21.0 | Similar to LMC |
| Host galaxy centered on quasar nucleus | — | $\gtrsim 18.6$ | Within radius of 2.8 kpc |

These cosmological parameters are used throughout the present paper. The redshift of the radio quasar PKS 2349−014 is $z = 0.173$; the apparent visual magnitude is $V = 15.3$ ($M_V = -23.4$); and the 6 cm and 11 cm fluxes are 0.7 Jy and 1.0 Jy, respectively (see Véron-Cetty & Véron 1993; Bolton & Ekers 1966; Searle & Bolton 1968; Wall 1972; and Pauliny-Toth & Kellermann 1972). No deep high-resolution VLA observations of PKS 2349−014 are available, although radio maps do exist which do not show the structure revealed by HST (cf. Antonucci 1985, Wills and Browne 1986). At the distance of PKS 2349−014, $1''$ is 1.9 kpc.

## 2. OBSERVATIONS

All of the figures presented in this paper were obtained from the same image, a 1400s exposure taken on September 17, 1994 with the Wide-Field Camera (WFC2) of HST (see Burrows 1994; Trauger et al. 1994; and Holtzman et al. 1995) using the wide-band visual filter, $F606W$. The scale of WFC2 is $0.0966''$ pixel$^{-1}$. Further details of the observational procedures are given in Paper II. The figures differ only in the scale or in the contrast at which the data is presented. For the purposes of this paper, the $V$ and $F606W$ photometric bands are sufficiently similar (see Paper II, Bahcall, Flynn, Gould, & Kirhakos 1994; and Holtzman et al. 1995) that we will use $V$ magnitudes and $F606W$ magnitudes interchangeably, although all HST-determined magnitudes are, strictly speaking, $F606W$ magnitudes. Since the main features of the images are apparent on the unprocessed HST exposures, we have chosen to present here the CCD data as they were obtained from the STScI, with the only further processing being the removal of cosmic rays with the aid of our shorter-exposure HST images (500s and 200s). In order to estimate the systematic errors for measured quantities, we have compared measurements made with and without subtracting a best-fit stellar PSF, measurements made with different size apertures, and measurements made with different length exposures.

Figure 1 highlights the most striking feature in the PKS 2349−014 images: a pair of thin, extended wisps. The wisps, which surround the quasar forming an almost complete ring, are smooth; no knots or bright spots are apparent. The wisps are brighter closer to the quasar, which is not located at the geometric center of the wisps. Given the visual similarity to other systems containing tidal wisps, for which there is satisfactory agreement between a theoretical model and observations, we suggest that PKS 2349−014 is undergoing a strong gravitational interaction.

We measured the surface brightness of the wisps at many different positions. The peak in the surface brightness is SW of the quasar and is about 21.4 mag arcsec$^{-2}$ (at approximately $0.3''$ south and $2.5''$ west of the quasar). The average surface brightness in the region covered by the wisps (roughly an annulus with radii $3.4''$ and $5.6''$) is about 22.2 mag arcsec$^{-2}$. For comparison, the surface brightness of the sky is about 22.1 mag arcsec$^{-2}$ in this image. The total extent of the major axis of the wisps, crudely represented as a ring, is about $10''$ (20 kpc), with a width, $2'' \pm 0.5''$ (4 kpc), that is difficult to measure precisely. The total visual magnitude of the region covered by the wisps is about $17.9 \pm 0.3$ (after subtracting the larger-scale estimated background light), which corresponds to an absolute visual magnitude of $M_V = -20.8$ (about 0.3 mag brighter than an $L^*$ field galaxy, cf. Kirshner et al. 1983; Efstathiou et al. 1988). The typical error in the surface brightness measurements is about 0.15 mag. The diffuse light between the wisps has an average surface brightness of $23.1 \pm 0.5$ mag arcsec$^{-2}$; the brightest region is closest to the quasar and the faintest region is most distant (and north) from the quasar (where the two separate wisps almost touch).

The faint, extended luminosity that is largely NE of the quasar is seen most clearly in the lower panel of Figure 2; this figure has been constructed by binning the light in $3 \times 3$ pixels into a single pixel. The average surface brightness in this component is 24.4 mag arcsec$^{-2}$. The major axis of this extended emission is $\sim 25''$, equivalent to 48 kpc, and the minor axis is $\sim 18''$, or 34 kpc. The apparent geometric center of the diffuse nebulosity is separated by about $2.7'' \pm 1''$ ($5 \pm 2$ kpc) from the center–of–light of the stellar quasar. The total magnitude of the extended emission (within the geometrical region defined above) is $18.0 \pm 0.3$, corresponding to an absolute magnitude of $M_V = -20.7 \pm 0.3$. Possible remnants of dust lanes can be seen to the northeast of the quasar nucleus.

In Figure 3, the contrast is much reduced, in order to show the distinct, faint galaxy that is located just above the quasar. In the following discussion, we assume that this galaxy is at the redshift of the quasar and is physically associated with PKS 2349−014. The galaxy is at a projected distance of $\sim 1.8''$, or 3.4 kpc, from the center of the quasar light. This close companion lies well within the region one would expect to be occupied by a host galaxy of PKS 2349−014.

The companion galaxy has an apparent magnitude of about $21.0 \pm 0.3$ (about 5.5 mag fainter than the quasar), which corresponds to a relatively faint galaxy, $M_V = -17.7 \pm 0.3$, at the redshift of PKS 2349−014. An expanded view of the central region of Figure 3, displayed in the insert, shows that the companion is about $1.0'' \pm 0.2''$ long, or 1.9 kpc in extent. The eastern wisp passes through the projected image of the companion galaxy.

Very close companion galaxies have also been detected in HST images of other nearby, luminous quasars (see, for example, the image of PKS 1302−102 in Paper II). However, we have not seen similarly striking evidence for interactions in any of the other previously-observed members of our luminous quasar sample.

Does PKS 2349−014 have a luminous host galaxy that is centered on the quasar nucleus? It is difficult to answer this question since the different components of diffuse light that are listed in Table 1 all overlap spatially. The visual light beyond $1.45''$ (15 pixels) around the quasar nucleus is dominated by the three prominent sources of diffuse light in the PKS 2349−014 system: the wisps, the extended nebulosity not centered on the quasar nucleus, and the faint companion galaxy. It is not clear whether or not the light within $1.4''$ is dominated by a host galaxy centered on the quasar nucleus. We can calculate a maximum brightness for the contribution of a host galaxy to the light within $1.45''$ of the quasar nucleus by summing up all of the light within this region, excluding the inner $0.5''$ (11% of the area) which is dominated by the quasar nucleus. We made the measurements by first subtracting a best-estimate of the nuclear component, using a measured PSF as described in Paper I and Paper II. We obtain $V(r \leq 1.45'') = 18.6$ and $M_V(r \leq 1.45'') = -20.1$.

The total amount of light measured within an annulus is a useful quantity since it is essentially independent of the resolution of the different components. We consider an annulus that was chosen to include all of the light from the wisps and the close companion and much of the light from the extended diffuse emission, but which excludes most of the light from the quasar nucleus. For an annulus with an inner radius of $0.5''$ and an outer radius of $9''$, we find $V = 16.8 \pm 0.1$ and $M_V = -21.9$, where the quoted uncertainty spans the range of values obtained on our three exposures (1400s, 500s, and 200s) analyzed with and without a point nucleus subtracted.

In addition to the very close companion which is featured in Figure 3, there are seven other galaxies around PKS 2349−014, marked A–G in Figure 1. These other galaxies have projected distances between 20 and 70 kpc from the quasar and have apparent magnitudes in the range of 19 to 22 ($M_V = -16.5$ to $M_V = -19.5$ if they are at the same distance as PKS 2349−014).

Table 2 contains information about the galaxies in the PKS 2349−014 field: apparent magnitude (aperture magnitudes measured with apertures $1''$ to $5''$, as appropriate, and typical uncertainties of $\pm 0.3$ mag), projected distance to the quasar and offsets in right ascension and declination from the center of the quasar light. The closest companion is listed as $L$ in the table.

Table 2: Galaxies in the PKS2349−014 Field

| | $m_{606W}$ | $d$ (arcsec) | $d$ (kpc) | $\Delta\,\alpha$ (arcsec) | $\Delta\,\delta$ (arcsec) |
|---|---|---|---|---|---|
| L | 21.0 | 1.8 | 3.4 | 1.8 | −0.4 |
| A | 21.4 | 11.2 | 21.3 | 8.3 | −7.6 |
| B | 18.8 | 15.9 | 30.2 | 14.4 | −6.8 |
| C | 21.4 | 17.4 | 33.1 | 4.9 | 16.7 |
| D | 21.0 | 20.4 | 38.7 | 17.4 | −10.6 |
| E | 20.9 | 27.8 | 52.8 | −26.5 | −8.2 |
| F | 22.1 | 33.0 | 62.7 | −28.0 | 17.4 |
| G | 19.4 | 35.2 | 66.9 | −6.5 | 34.6 |

## 3. DISCUSSION

The HST view of PKS 2349−014 is similar to the images of a number of interacting galaxies in the Atlas of Peculiar Galaxies (Arp 1966), especially to Arp 224, which has a redshift of $z = 0.020$. Mazzarella & Boroson (1993) present a particularly revealing image of Arp 224, which shows wisps and faint diffuse emission analogous to PKS 2349−014 and even has similar spatial dimensions, i. e., for Arp 224 the wisps are $\sim 25$ kpc in length and the extended nebulosity is $\sim 45$ kpc. [According to Stauffer (1982), Arp 224 has an active nucleus with low ionization emission (LINER), while Mazzarella & Boroson (1993) classified the nuclear emission as being caused by starbursts.]

It seems plausible that the same mechanism that produces the pyrotechnics in Arp 224 may also be responsible for igniting PKS 2349−014. In this scenario, the galaxy interactions would result in an increased supply of fuel to a massive black hole. It would be very interesting to make a detailed dynamical simulation of the interactions that gave rise to the curved wisps and other observed features in the PKS 2349−014 system.

The large, faint diffuse emission seen most clearly in the lower panel of Figure 2 has an average surface brightness of 24.4 mag arcsec$^{-2}$ (about 12% of the sky brightness) and a total absolute magnitude of $M_V = -20.7$, about 0.2 mag fainter than an $L^*$ galaxy. Thus even relatively low surface brightness objects can be detected in moderately long exposures taken with the WFC2 through a broad-band filter.

The detection of the large, low surface brightness emission in the PKS 2349−014 system provides additional evidence that the absence of detectable diffuse emission around other quasars discussed in Paper I and Paper II indicates that the host galaxies of those other quasars are not very luminous. The instrument (WFC2), the filter (F606W), the exposure time (typically 1100s), and the brightness of the quasar nucleus were similar for seven of the previous eight cases (3C 273 was somewhat different). Moreover, none of the previously-analyzed quasars in our sample are interacting with such an obvious diffuse environment like that of PKS 2349−014, indicating that the regions in which different luminous quasars exist are not all the same.

For PKS 2349−014, there is a companion galaxy at a projected distance of about 3 kpc of the quasar center-of-light. The projected distance from the companion galaxy to the center of the quasar light is (for $H_0 = 100$ km s$^{-1}$ Mpc$^{-1}$) less than one half the distance between our sun and the center of the Galaxy. Furthermore, the PKS 2349−014 companion has a size and brightness similar to the Large Magellanic Cloud. The Large Magellanic Cloud half-light-radius is $3.0°$ (Bothun & Thompson 1988). If the LMC were at the distance of PKS 2349−014, the LMC would have an apparent half-light size of $1.4''$, comparable to what is observed, $1.0''$, for the close companion of PKS 2349−014. The estimated absolute visual magnitude of the PKS 2349−014 companion is $−17.7$; the absolute half-light visual magnitude of the LMC is $−17.6$ (Bothun & Thompson 1988). The companion of PKS 2349−014 would be expected to merge with the quasar in $\sim 10^7$ years due to dynamical friction, if there is stellar material in the quasar nucleus-companion galaxy system comparable to what is present in the Galaxy (cf. Tremaine 1976).

Seven other galaxies with apparent visual magnitudes in the range 19 to 22 are within 70 kpc of the center of light of PKS 2349−014. Using the observed density of galaxies in the central regions of the other two WFC2 CCDs, we estimate that less than three galaxies would be expected by chance in the projected area corresponding to a projected distance of 70 kpc from PKS 2349−014. Obtaining the redshifts of the eight companion galaxies listed in Table 2 would permit more accurate modeling of the PKS 2349−014 system.

Since both the wisps and the extended diffuse blob appear to envelop the center of the quasar light, we are limited in what we can say about the brightness of a possible host galaxy that might be centered on the quasar. Within an annulus centered on the quasar and having inner and outer radii of 0.9 kpc ($0.5''$) and 2.75 kpc ($1.45''$), the total amount of light measured is $V = 18.6$ and $M_V = −20.1$. We are unable to determine if most of this light comes from a host galaxy centered on the quasar or is due to diffuse emission between the wisps. For a much larger annulus (that includes the wisps, the companion galaxy, and much of the extended diffuse emission) with inner and outer radii of 0.9 kpc and 17.1 kpc ($9.0''$), the total amount of light measured is $V = 17.0$ and $M_V = −21.7$.

The previous ground-based observations of PKS 2349−014 have lower angular resolution than the HST images and therefore do not provide details of the light distribution of the different diffuse components. The ground-based observations have been analyzed by assuming that there is a single diffuse component centered on the stellar light of PKS 2349−014. Boyce, Phillips, & Davies (1993) found $M_V(\text{host galaxy}) = −22.75 \pm 0.1$ and Véron-Cetty & Woltjer (1990) found $M_V(\text{host galaxy}) = −22.6$ as the best-estimate of the host galaxy luminosity based upon fits to a de Vaucouleurs profile centered on the quasar nucleus. These estimated host galaxy magnitudes exceed the luminosity of the brightest galaxies in rich clusters, which typically have $M_V = -22.0$ (see Hoessel and Schneider 1985; Postman and Lauer 1995). Dunlop et al. (1993), using $K$-band images, reported the presence of a luminous host galaxy and evidence for a strong interaction in the PKS 2349−014 system; they determined a ratio of 2.7 for the luminosity of the quasar nucleus ($M_V = -23.4$) to the luminosity of the host galaxy. The various ground-based observations suggested a characteristic scale length for the observed light in the range $1.5''$ to $3''$.

The previous ground-based results gave brighter luminosities for a host galaxy centered on the quasar than is indicated by the HST observations. The reason for this discrepancy is apparently the complex, multiple-component nature of the diffuse light in the PKS 2349−014 system. Guided by the higher-resolution images obtained from HST, we have represented the system by four components of diffuse light and a (stellar) quasar nucleus (see Table 1). If, following the same procedure as was used in analyzing the ground-based observations, we change the model and represent all of the diffuse light by a single, featureless de Vaucouleurs profile and fit to an azimuthal average of the observed light distribution, we obtain $M_V = -22.4$. Thus the host galaxy luminosity obtained by fitting the (multiple-component) HST image to a single de Vaucouleurs profile centered on the quasar is in good agreement with previous ground-based results. However, the effective radius of the de Vaucouleurs profile that best fits the HST images is $6''$, which is considerably larger than the values that were obtained with the ground-based data.

In summary, the excellent angular resolution of the HST has revealed that the luminous quasar PKS 2349−014 is a complex system of apparently interacting components. The thin curved wisps shown in Figure 1 and Figure 3 are characteristic signals for tidal interactions among galaxies. The compressed image (see the lower panel of Figure 2) makes obvious the presence of an extended, low surface brightness nebulosity that is displaced from the quasar nucleus by about 5 kpc. This low surface brightness nebulosity could be a galaxy that is interacting with the quasar or perhaps tidal debris from the interaction. The close companion, the LMC look-alike highlighted in the insert to Figure 3, appears to be so close to the quasar nucleus that it seems likely to be gobbled up in the near future.

What causes the quasar phenomenon in PKS 2349−014? Many different questions that require futher study, theoretical and observational, are suggested by Figure 1-Figure 3. For example, does numerical modeling suggest that the different components are most likely due to tidal interactions among ordinary galaxies? Or, is PKS 2349−014 a ring galaxy that has been punctured by a massive black hole? What is the role of the companion LMC-like object? What is its redshift? What are the kinematics of the other diffuse components? What is the nature of the extended nebulosity? How much of the emission in each of the observed components is emission lines (O[III]$\lambda$5007 and H$\beta$ are in the $F606W$ bandpass) and how much is starlight? Will high resolution radio maps show structure that reflects some or all of the optical components? It seems likely that additional observations, when combined with detailed theoretical modeling, will provide important insights into how luminous quasars are fueled.

We are grateful to R. Blandford, T. Boroson, P. Goldreich, A. Gould, J. Halpern, B. Jannuzi, J. MacKenty, J. Ostriker, S. Phinney, D. Richstone, M. Schmidt, D. Spergel, L. Spitzer, A. Stockton, M. Strauss, A. Toomre, and B. Wills for valuable discussions, comments, and suggestions. D. Saxe provided expert help with the figures. We would like to thank Digital Equipment Corporation for providing the DEC4000 AXP model 610 system used for the computationally-intensive parts of this project. This work was supported in part by NASA contract NAG5-1618 and grant number GO-5343 from the Space Telescope Science Institute, which is operated by the Association of Universities for Research in Astronomy, Incorporated, under NASA contract NAS5-26555.

## FIGURE CAPTIONS

Fig. 1.— The environment of PKS 2349−014. This figure shows PKS 2349−014 and its galaxy environment. Thin wisps (possibly tidal in origin) indicate that diffuse material near the quasar is being strongly affected by gravitational interactions. The candidate companion galaxies labeled "A–G" are all within a projected distance of 70 kpc from the center of light of the quasar. The image was obtained with a 1400 s exposure using the HST WFC2 and the $F606W$ filter.

Fig. 2.— The components of PKS 2349−014. The top panel shows a schematic diagram, not to scale, of the four components of PKS 2349−014 that are clearly visible on the HST images plus a faint host galaxy which, if present, is hidden by the quasar nucleus and other prominent components. The bottom panel gives a clear view of the faint offset nebulosity that envelopes PKS 2349−014; this panel was obtained by binning $3 \times 3$ pixels into a single pixel.

Fig. 3.— The wisps and the LMC-like companion. This figure (which is the same as Figure 1 except that it is reproduced at a much reduced contrast) shows clearly the wisps of PKS 2349−014, which approximately surround the quasar, and the presence of a small companion galaxy 1.8″ from the quasar. The inset shows a close-up of PKS 2349−014 that highlights the close companion galaxy.

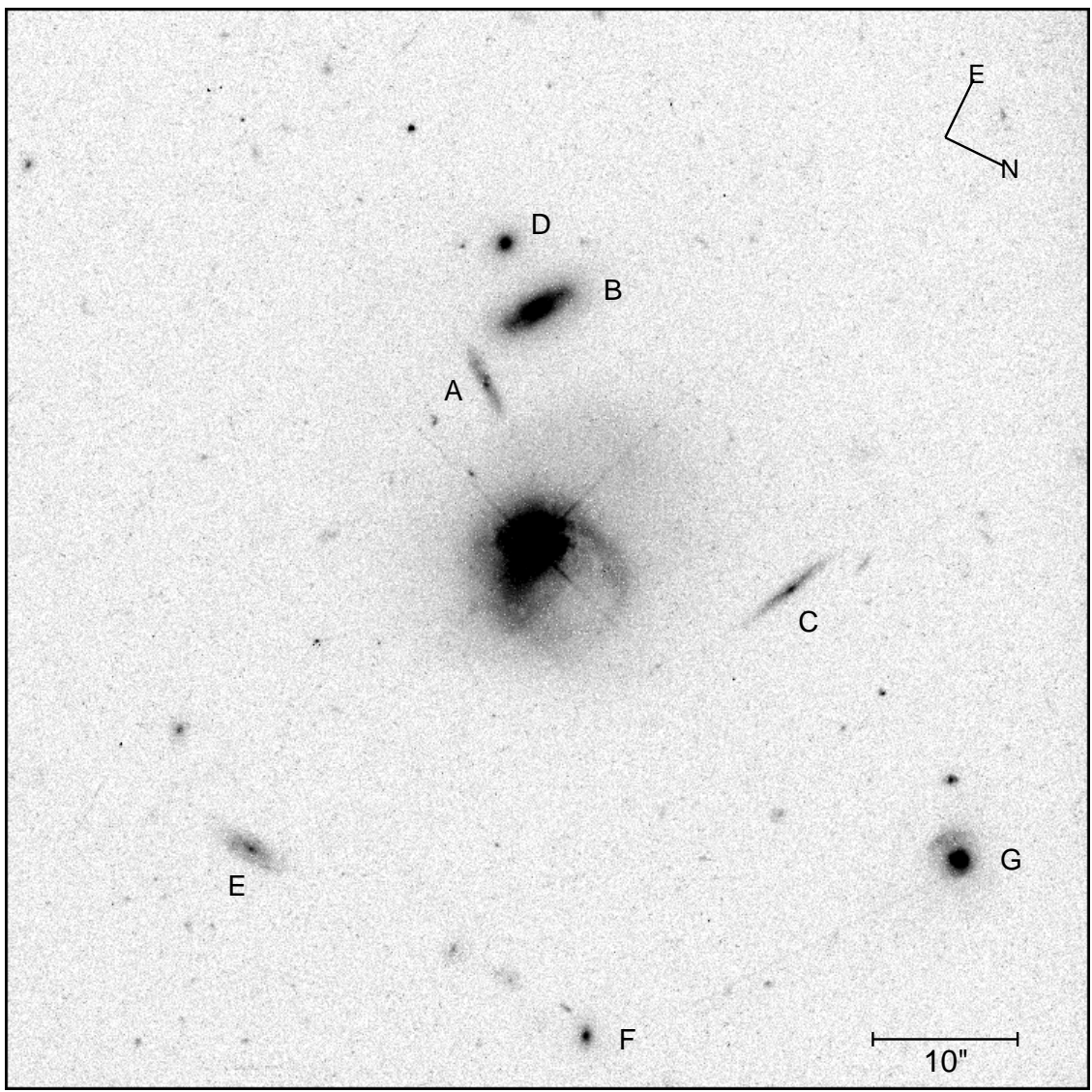

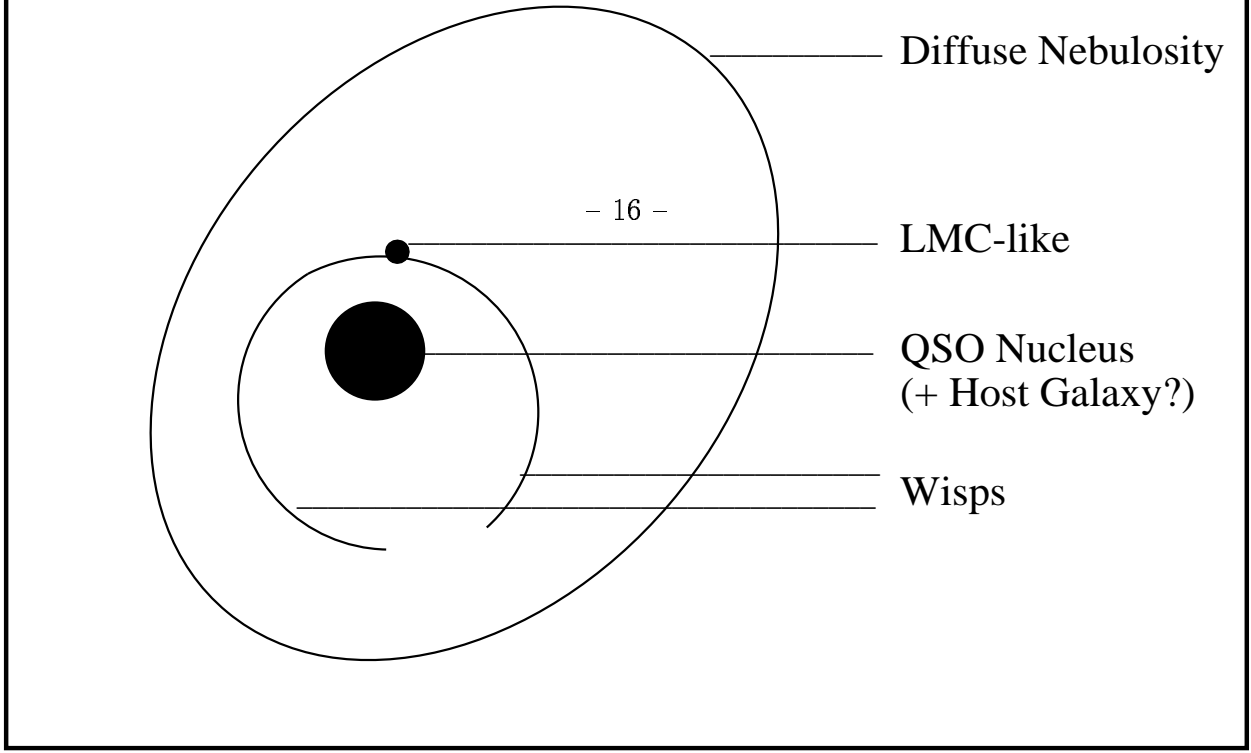

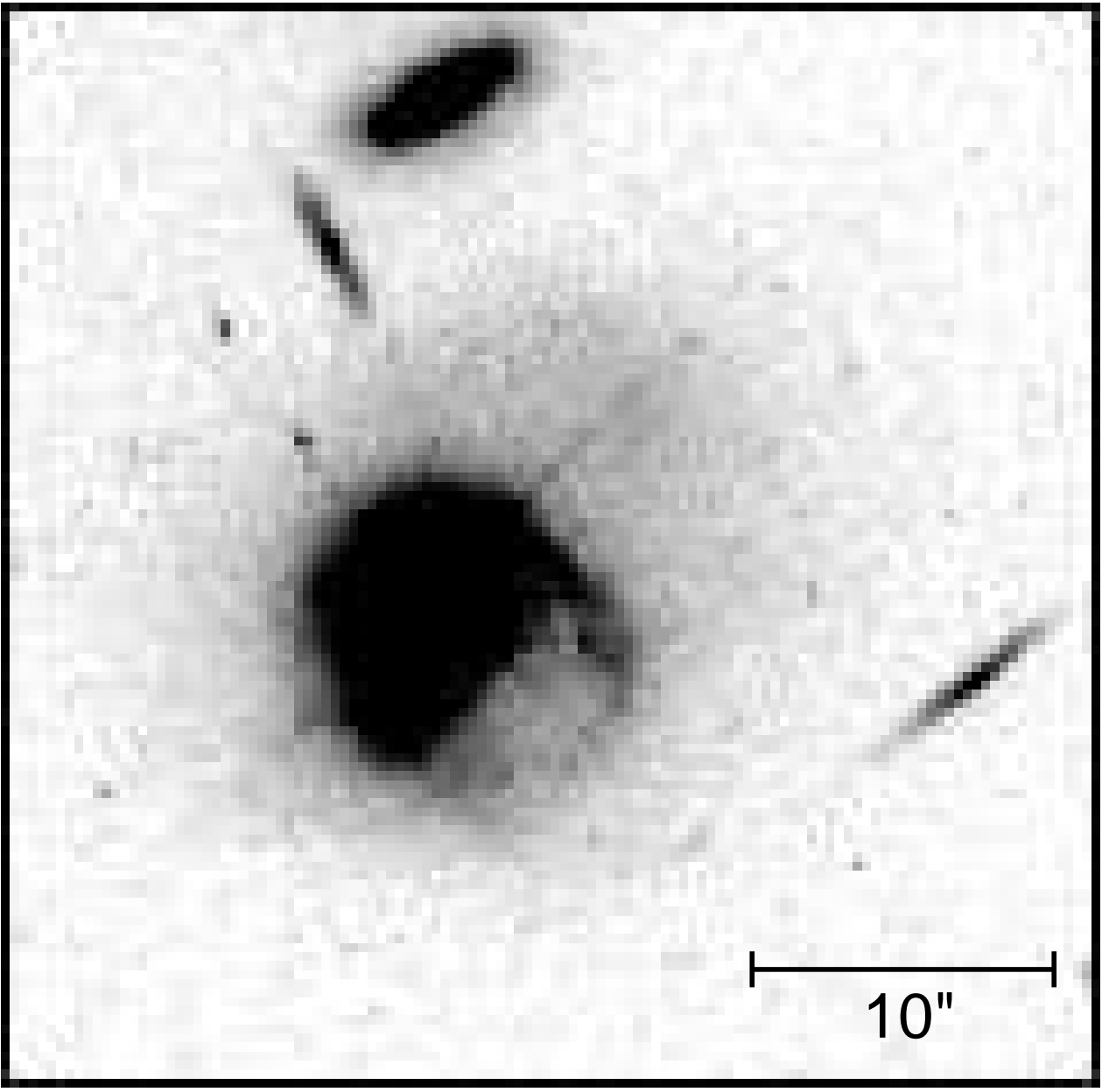

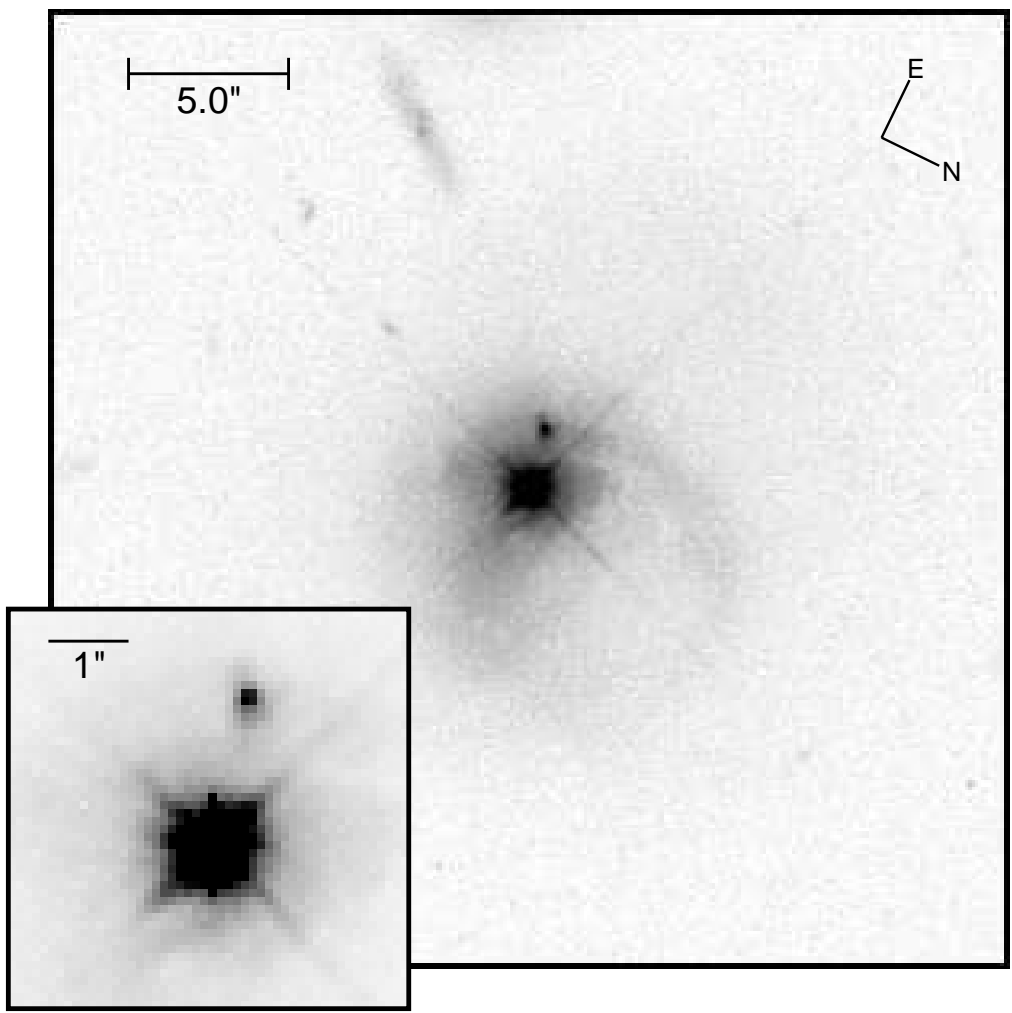